\documentclass[aps,prl,letterpaper,twocolumn,showpacs]{revtex4}
\usepackage{amsmath,bm}
\usepackage{amssymb}
\usepackage{graphicx}
\usepackage{multirow}
\begin{document}
%\linenumbers

\title{Polarization-dependence of palladium deposition on
ferroelectric lithium niobate (0001) surfaces}
\author{Seungchul Kim}
\affiliation{The Makineni Theoretical Laboratories, Department of
Chemistry, University of Pennsylvania, Philadelphia, Pennsylvania
19104--6323, USA}
\author{Michael Rutenberg Schoenberg}
\affiliation{The Makineni Theoretical Laboratories, Department of
Chemistry, University of Pennsylvania, Philadelphia, Pennsylvania
19104--6323, USA}
\author{Andrew M. Rappe}
\email[corresponding author:\ ]{rappe@sas.upenn.edu}
\affiliation{The Makineni Theoretical Laboratories, Department of
Chemistry, University of Pennsylvania, Philadelphia, Pennsylvania
19104--6323, USA}
\date{\today }

\begin{abstract}

We investigate the effect of ferroelectric polarization direction
on the geometric properties of Pd deposited on the positive and
negative surfaces of LiNbO$_3$ (0001).  We predict preferred
geometries and diffusion properties of small Pd clusters using
density functional theory, and use these calculations as the basis
for kinetic Monte Carlo simulations of Pd deposition on a larger
scale. Our results show that on the positive surface, Pd atoms
favor a clustered configuration, while on the negative surface, Pd
atoms are adsorbed in a more dispersed pattern due to suppression
of diffusion and agglomeration. This suggests that the effect of
LiNbO$_3$ polarization direction on the catalytic activity of Pd
[J. Phys. Chem. \textbf{88}, 1148 (1984)] is due, at least in
part, to differences in adsorption geometry. Further
investigations using these methods can aid the search for
catalysts whose activities switch reversibly with the polarization
of their ferroelectric substrates.
\end{abstract}


%77.84.Ek  Niobates and tantalates
%68.43.-h  Chemisorption/physisorption: adsorbates on surfaces
%68.43.Bc  Ab initio calculations of adsorbate structure and reactions
%68.43.Fg  Adsorbate structure (binding sites, geometry)
%68.43.Hn  Structure of assemblies of adsorbates
%          (two- and three-dimensional clustering)
%68.43.Jk  Diffusion of adsorbates, kinetics of coarsening and aggregation
%68.43.Mn  Adsorption kinetics
%68.35.Md  Surface thermodynamics, surface energies
%82.65.+r  Surface and interface chemistry;
%          heterogeneous catalysis at surfaces

\maketitle

The spontaneous polarization of ferroelectric materials has
enabled their use in technological devices ranging from SONAR to
random access memory.  While most current applications of
ferroelectrics result primarily from their bulk properties,
polarization can also impact the surface properties of these
materials, including surface stoichiometry~\cite{Noguera00p367,
Levchenko08p256101, Yun07p4636},
geometry~\cite{Levchenko08p256101, Yun07p4636}, and electronic
structure~\cite{Noguera00p367, Inoue85p2827, Park00p4418,
Yun07p4636}. These polarization dependent differences in intrinsic
surface properties also affect their interactions with
adsorbates,~\cite{Yun07p15684, Yun07p13951, Kolpak07p166101,
Li08p473, Zhao09p1337, Inoue84p1148, Inoue92p2222, Saito02p10179}
as evidenced by differences in adsorption
energies~\cite{Yun07p15684, Yun07p13951, Kolpak07p166101,
Li08p473, Zhao09p1337} or the rates of surface catalyzed
reactions~\cite{Inoue84p1148, Inoue92p2222, Saito02p10179}.
Polarization orientation can, in turn, affect the chemical
properties of the adsorbates themselves. For example, Inoue
\emph{et al.}~\cite{Inoue84p1148} showed that the activation
barrier for CO oxidation by Pd adsorbed on a LiNbO$_3$ surface
changes by 30~kJ/mol, depending on polarization orientation. The
notion that a ferroelectric substrate's polarization can affect
the activity of a supported catalyst suggests the intriguing
possibility that the activity of a catalyst could be modulated
reversibly by switching the polarization of a ferroelectric
substrate.

Despite ample evidence showing that metals on oppositely poled
ferroelectric surfaces have different catalytic properties, the
mechanism underlying these differences is not well understood. The
most prevalent explanation for this phenomenon has been that the
difference in charge between the two surfaces alters the
electronic properties of the catalyst~\cite{Inoue84p1148,
Inoue92p2222, Inoue85p2827, Saito02p10179}. However, this
explanation may be incomplete, as metal adsorption geometries
significantly impact their catalytic
properties~\cite{Norskov08p2163}. Since ferroelectric surfaces may
have different geometries depending on the sign of their
polarization~\cite{Noguera00p367, Levchenko08p256101, Yun07p4636},
it is plausible that the geometries of metals adsorbed onto them
also differ~\cite{Zhao09p1337, Yun09p3145}. Two recent studies of
Pd adsorption on oppositely poled LiNbO$_3$ (0001) surfaces
present conflicting conclusions regarding the possible effect of
polarization on metal adsorption geometry. Yun \emph{et
al.}~\cite{Yun09p3145} showed that large Pd clusters form on both
the positive ($c^+$) and negative ($c^-$) surfaces of LiNbO$_3$,
suggesting that polarization has little impact on metal adsorption
geometry. In contrast, Zhao \emph{et al.}~\cite{Zhao09p1337}
observed large Pd clusters only on the $c^+$ surface, and a more
planar geometry on the $c^-$ surface, suggesting that polarization
strongly affects metal adsorption geometry. In addition, only Zhao
\emph{et al.} observed a difference in CO temperature programmed
deposition between the two surfaces, suggesting that when
polarization affects the activity of adsorbed catalysts, it does
so, at least in part, by altering their geometries.

In view of the ambiguity of experimental investigations of the
relationship between polarization and metal adsorption geometry,
we address this question on a microscopic level using theoretical
methods. Here we investigate the energetics and kinetics of Pd
deposition on LiNbO$_3$ (0001) surfaces using a combination of
density functional theory (DFT) calculations and kinetic Monte
Carlo (KMC) simulations. We first calculate the adsorption
geometries of clusters on the $c^+$ and $c^-$ surfaces. We then
model the range of possible diffusion and agglomeration processes
of these clusters using the nudged elastic band (NEB)
method~\cite{Henkelman00p9901}. Finally, we use the activation
barriers of these processes as inputs for a KMC simulation of the
deposition of Pd on LiNbO$_3$ on a larger scale. To our knowledge,
this is the first theoretical study of metal adsorption kinetics
on a ferroelectric surface.

\begin{figure}[t]
  \centering
  \includegraphics[width=0.45\textwidth]{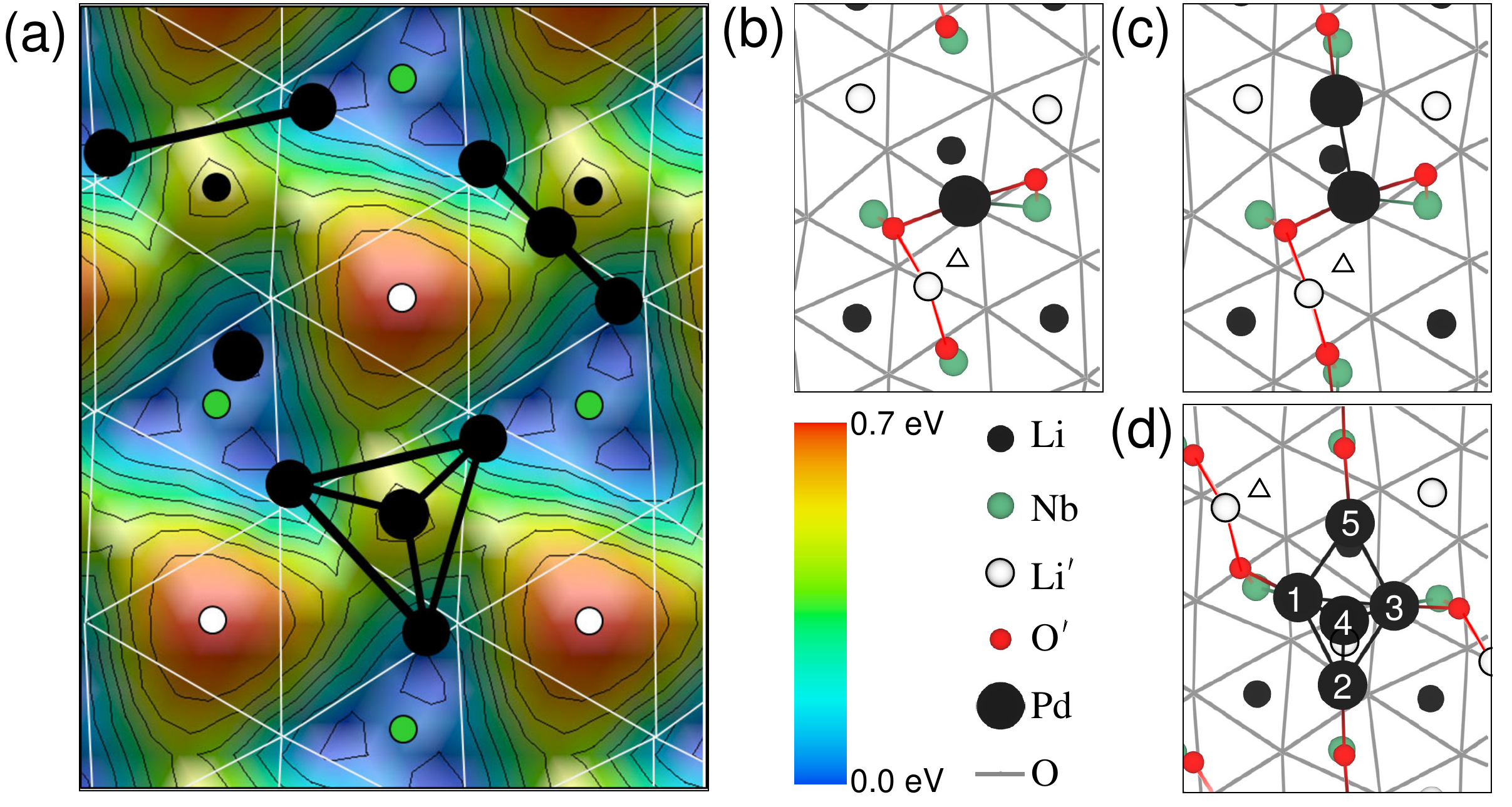}
  \caption{(Color online)  (a) Left Side: Monomeric potential
energy surface and minimum energy geometries of Pd$_1$, Pd$_2$,
Pd$_3$ and Pd$_4$ on $c^+$. Contour spacing is 0.1 eV. Right side:
Minimum energy geometries of (b) Pd$_1$, (c) Pd$_2$, and (d)
Pd$_5$ on $c^-$. Triangles are original positions of Li$'$. The
geometries for Pd$_3$ and Pd$_4$ are similar to those of atoms
1-2-3 and 1-2-3-4 in (d).}
 \label{f.geo}
\end{figure}

In all DFT calculations, we studied five trilayer thick LiNbO$_3$
slabs with $\sqrt{3}\times\sqrt{3}$ surface supercells
(Supplementary Fig.\ S1). We passivated surface charges with one
Li atom per primitive supercell on the $c^+$ surface, and one O
and one Li atom on $c^-$, in accordance with the findings of
Levchenko and Rappe that this is the most thermodynamically stable
surface composition for LiNbO$_3$~\cite{Levchenko08p256101}.
Hereafter, we denote these passivation atoms as Li$'$ and O$'$. To
remove spurious interactions between slabs in different
supercells, we separated the slabs with more than 13~\AA\ of
vacuum and employed a dipole correction~\cite{Bengtsson99p12301}.
DFT total energy calculations~\cite{Ihm79p4409} were performed
using the generalized gradient approximation
(GGA-PBE)~\cite{Perdew96p3865} and implemented using the Quantum
Espresso package~\cite{Giannozzi09p395502}. Atoms were represented
using norm-conserving nonlocal
pseudopotentials~\cite{Rappe90p1227,Ramer99p12471}  generated
using the OPIUM code~\cite{Opium}. We allowed relaxation of Pd
atoms and the first two layers of the LiNbO$_3$ surface, but kept
the remaining portion of the slab fixed.

Our DFT results for Pd adsorption show differences in preferred
binding geometries and diffusion barriers between the $c^+$ and
$c^-$  surfaces. On the $c^+$ surface, Pd adsorption minimally
changes the geometry of the LiNbO$_3$ surface itself. As a result,
the monomeric potential energy surface (PES, Fig.\ \ref{f.geo}(a))
essentially determines the geometries of multiple Pd atoms
adsorbed on this surface. In particular, all Pd atoms that we
predict to bind directly to the $c^+$ surface prefer sites close
to monomeric potential energy minima. However, because the
distances between minima do not match well with optimal Pd-Pd bond
lengths, binding of large numbers of Pd atoms both to the $c^+$
surface and to each other is unfavorable. For example, adsorption
geometries of three and four Pd atoms both include one atom that
bonds only to other Pd atoms and does not interact directly with
the surface (Fig.\ \ref{f.geo}(a)). Further, we found no
metastable planar structure of four atoms, suggesting that the Pd
agglomeration barrier on the $c^+$ surface is negligible.

\begin{figure}[t]
  \centering
  \includegraphics[angle=0, width=0.45\textwidth]{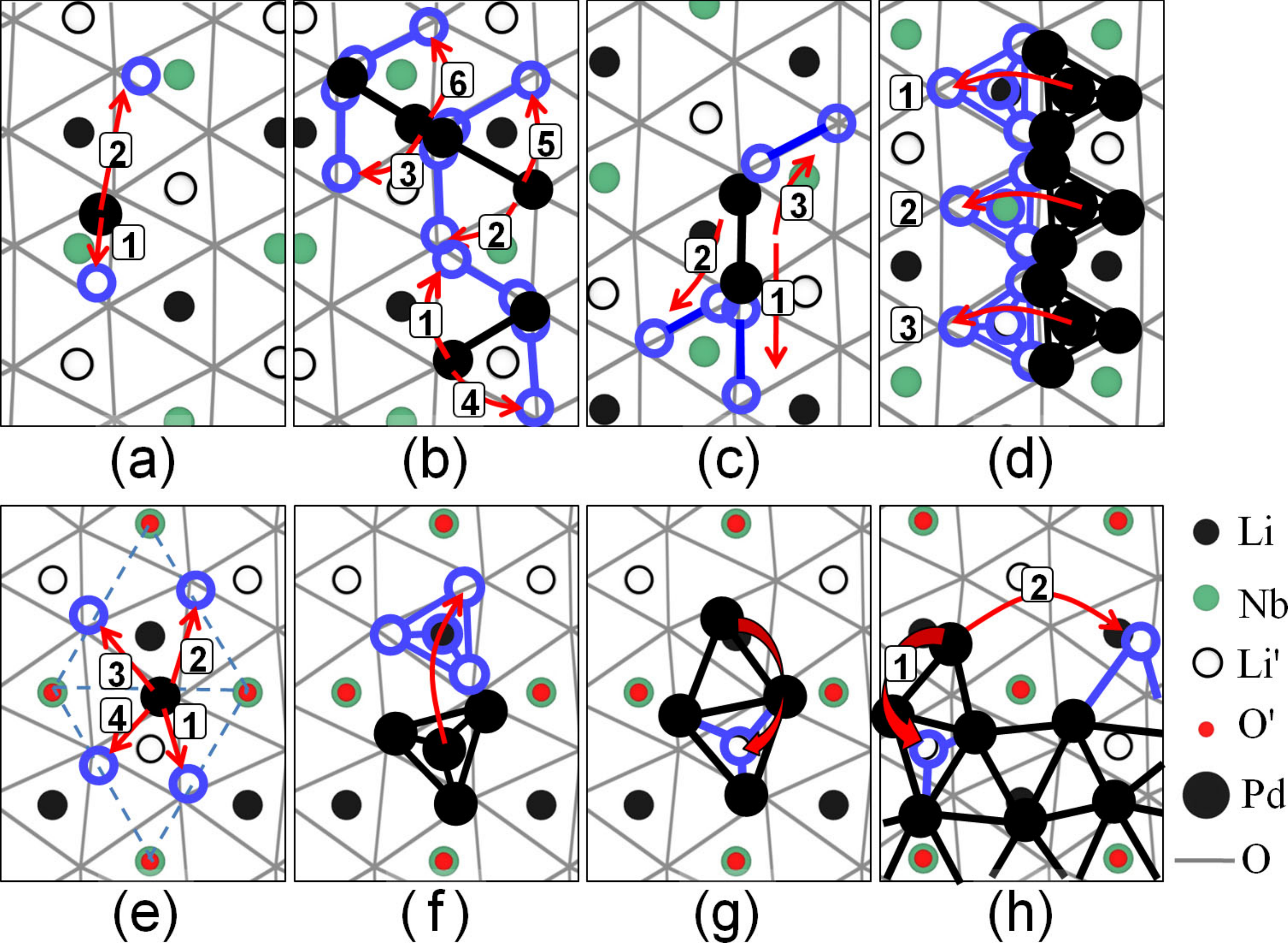}
\caption{(Color online) Schematic drawings of kinetic events.
Top:  Diffusion events on $c^+$. (a) Pd$_1$ hopping, (b) Pd$_2$
walking, (c) Pd$_2$ sliding and (d) Pd$_4$ rolling. Bottom:
Diffusion  events on $c^-$. (e) Pd$_1$ hopping, (f) Pd$_4$
rolling, (g) Pd$_4$ agglomeration, and (h) agglomeration at large
cluster (path 1) and in-plane motion (path 2). Large black circles
and open blue circles denote initial and final positions for
diffusion events. after the events. Blue dashed lines in (e)
indicate that the monomer hops between O$'$ bridge sites. }
  \label{f.neb}
\end{figure}

Diffusion and agglomeration processes on the $c^+$ surface are
also impacted by the monomeric PES.  In the monomeric case, this
relationship is direct, as the two unique paths for monomer
hopping (Fig.\ \ref{f.neb}(a)) are the only paths containing
saddle points on the PES. Because the diffusion processes of
clusters of up to four atoms on the $c^+$ surface are dominated by
the movement of a single atom, they have diffusion barriers
similar to monomer hopping, a fact that can also be explained by
the monomeric PES. For example, in dimer walking, a process in
which one atom in a dimer steps between two potential minima and
stays bonded to the other Pd, paths 2 and 3 have activation
barriers similar to monomer hopping paths 1 and 2, respectively
(Fig.\ \ref{f.neb}(b)). Similarly, though dimer sliding involves
movement of both Pd atoms, one stays adjacent to the same Nb atom
and thus, its movement contributes little to the activation
barrier (Fig.\ \ref{f.neb}(c)). Finally, because tetramer rolling
requires movement of only one Pd atom out of a monomeric potential
well, it too has an activation barrier similar to that of monomer
hopping. The combination of low Pd diffusion barriers and a
negligible agglomeration barrier suggests that formation of large
Pd clusters is favorable on the $c^+$ surface.

In contrast to its behavior on the $c^+$ surface, Pd adsorption on
the $c^-$ surface substantially alters the geometry of the surface
itself.  This is largely due to interactions with O$'$ atoms,
which along with Li$'$ atoms terminate the $c^-$ surface for
charge passivation~\cite{Levchenko08p256101}.  The formation of
Pd-O$'$ bonds makes adsorption onto the $c^-$ surface much more
favorable than on $c^+$ (Supplementary Table\ S1).  However this
also leads to complex adsorption geometries, as the O$'$ atom is
free to tilt its bond to Nb substantially in order to accommodate
bonding to Pd. Despite this geometric complexity, we find that the
Pd adsorption geometries on the $c^-$ surface can be understood in
the context of maximizing the number of Pd-Pd and Pd-O$'$ bonds
formed. For example, this explains why the 2D planar (not shown)
and 3D configurations of clusters of four or five Pd atoms on the
$c^-$ surface, which have identical numbers of Pd-Pd/Pd-O$'$
bonds, have binding energies within 0.1 eV of one another.

The inherent assumption in this analysis, that Pd-O$'$ and Pd-Pd
bonds have similar strengths, is justified by two facts. First,
the average energies of these bonds, defined as adsorption
energies divided by the number of Pd-Pd and Pd-O$'$ bonds, in
clusters of 1--5 Pd atoms are uniform (1.03 $\pm$ 0.05 eV).
Second, these average bond energy values are similar to bond
strengths in free clusters of 2--5 Pd atoms (1.06 $\pm$ 0.08 eV).
From here on, we denote Pd-Pd and Pd-O$'$ bonds as Pd-X.

\begin{table}
  \begin{ruledtabular}
  \begin{tabular}{cccccccc}
   & path           & 1    & 2    & 3    & 4    & 5    & 6    \\
   \hline
   $c^+$     & Pd$_1$ hopping & 0.09 & 0.39 &      &      &      &      \\
        & Pd$_2$ walking & 0.26 & 0.12 & 0.38 & 0.24 & 0.38 & 0.17  \\
        &                & 0.32 & 0.02 & 0.41 & 0.24 & 0.38 & 0.17  \\
        & Pd$_2$ sliding & 0.35 & 0.36 & 0.30 &      &      &      \\
        &                & 0.32 & 0.33 & 0.27 &      &      &      \\
        & Pd$_3$ walking & 0.23 & 0.19 & 0.27 & 0.24 & 0.11 & 0.18  \\
        &                & 0.12 & 0.11 & 0.46 & 0.24 & 0.11 & 0.18  \\
        & Pd$_4$ rolling &$<$0.01& 0.21& 0.06 &      &      &      \\
        &                & 0.40 & 0.14 & 0.13 &      &      &      \\
        & Pd$_2$ dissociation & 0.55 \\
   \hline
   $c^-$& Pd$_1$ hopping & 0.82 & 0.87 & 0.87 & 0.82 &      &      \\
        & Pd$_2$ sliding & 1.06 &                                  \\
        & Pd$_3$ flipping (Li$'\rightarrow$Li) & 1.30 & (1.31) & & & \\
        & Pd$_4$ rolling  (Li$'\rightarrow$Li) & 0.85 & (0.70) \\
        & Pd$_2$ dissociation & 0.77 \\
        & large cluster, in-plane & 1.12 \\
        & large cluster, agglom.  & 0.74 & (0.93) \\
        & Pd$_4$, agglom.   & 0.59 & (0.61) \\
  \end{tabular}
  \end{ruledtabular}
\caption{Diffusion activation barriers (in eV) of processes drawn
in Fig.\ \ref{f.neb} and described in the text. Barriers of
reverse processes are written below forward processes ($c^+$) or
in parentheses ($c^-$).}
  \label{t.Etr}
\end{table}

Just as Pd-X bonds are the primary contributors to the favorable
adsorption energy of Pd clusters on the $c^-$ surface, breaking
these bonds, particularly Pd-O$'$, is the primary barrier to
cluster diffusion and agglomeration. Based on the argument above
that Pd-X bonds have similar energies, we would expect the
activation barriers for processes that require one Pd-O$'$ bond
breaking to be approximately 1~eV. However, actual activation
barriers are lower, because other atoms can move to more favorable
positions in the process of transition state formation. This is
especially true when new Pd-X bonds are formed before a Pd-X bond
is fully broken. Consistent with our understanding of the effect
of Pd-X bond breaking on the activation barriers of diffusion
processes, we find that in-plane movement at the boundary of a
cluster has a high activation barrier (path 2 in Fig.\
\ref{f.neb}(h)), because it requires breaking of two Pd-X bonds.

Our DFT calculations show that diffusion barriers are
substantially lower on the $c^+$ surface ($\lesssim$ 0.4 eV) than
on the $c^-$ surface ($\gtrsim$ 0.8 eV). These activation
energies, $E_{a}$, can be compared in a physically meaningful way
when converted to expected event-event time intervals, $\tau$,
using Arrhenius kinetics~\cite{Brune98p121}, $\frac{1}{\tau} = \nu
e^{-E_{a}/k_{B}T}$. Assuming an attempt frequency
($\nu=10^{12}$~sec$^{-1}$~\cite{Xu06p1351}) the time scales of
diffusion events on the $c^+$ and $c^-$ surfaces are on the order
of microseconds and minutes, respectively.  We can then infer that
at a deposition rate of
$\approx$~0.01--0.1~ML/s~\cite{Yun09p3145}, on average, each new
monomer deposited on the $c^+$ surface will aggregate to an
existing cluster before the next atom is deposited. In contrast,
we would expect many Pd atoms to be deposited in the vicinity of a
given Pd atom on the $c^-$ surface between diffusion events of
that atom. We thus infer that Pd atoms will agglomerate into much
larger clusters on the $c^+$ surface than on the $c^-$ surface of
LiNbO$_3$.

We conducted kinetic Monte Carlo (KMC) simulations in order to
characterize this inferred difference in adsorption geometries.
Onto a 10~nm $\times$ 10~nm surface, we randomly deposited Pd
atoms one by one at a selected deposition rate. We then allowed Pd
atoms and clusters to attempt a series of diffusion and
agglomeration processes, with probabilities based on our
calculated activation barriers, at a constant attempt frequency of
$10^{12}$ sec$^{-1}$~\cite{Xu07p3133} for all events (Full KMC
description in Supplement).

\begin{figure}[t]
  \centering
  \includegraphics[angle=0, width=0.4\textwidth]{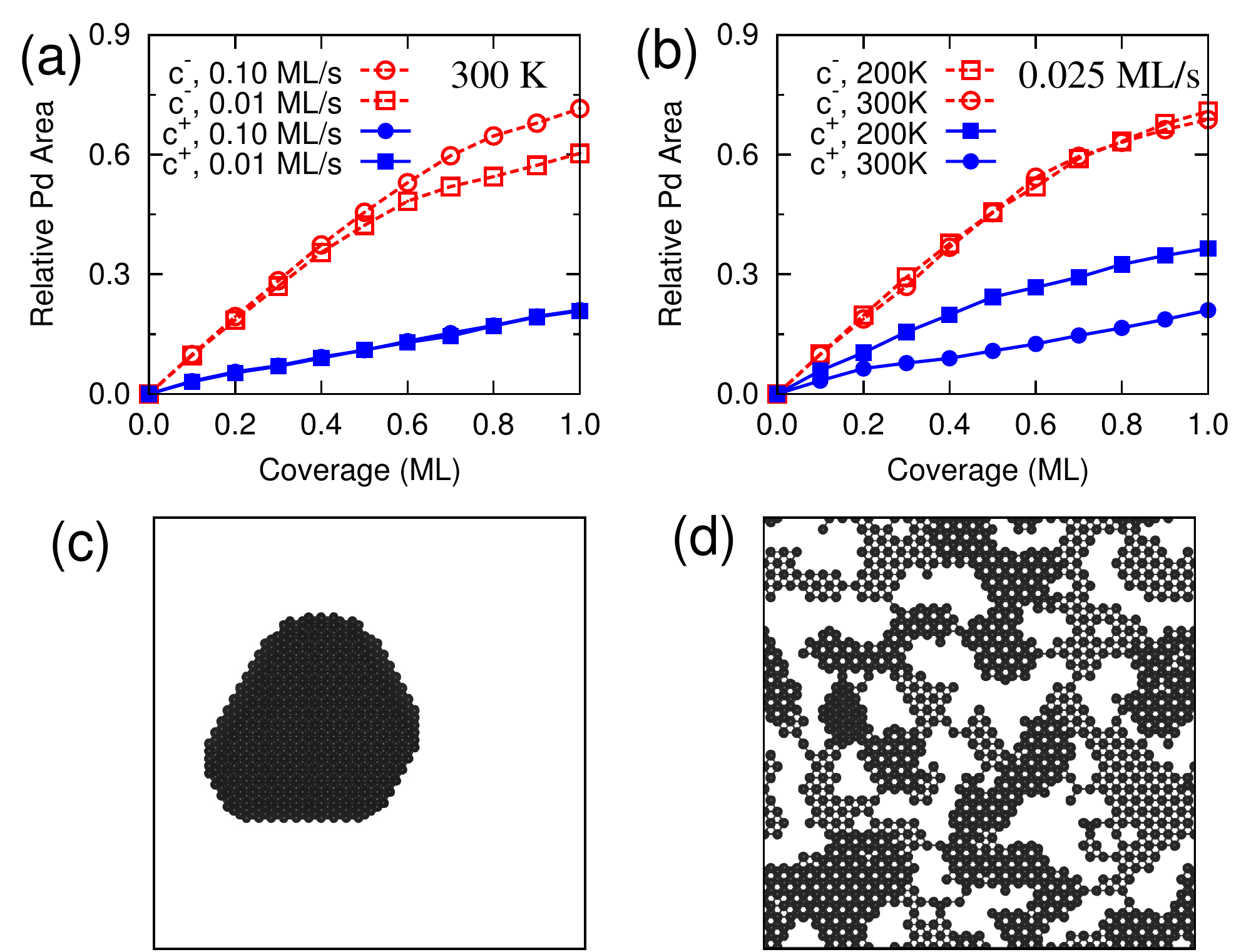}
\caption{(Color online) Proportion of surface area covered by Pd
for simulations at a range of (a) deposition rates at 300~K and
(b) temperatures with a deposition rate of 0.025~ML/s. KMC
snapshots of 1.0~ML on (c) $c^+$ and (d) $c^-$ at 300~K with a
0.025~ML/s deposition rate.}
 \label{f.kmc}
\end{figure}

Our KMC simulations confirm that Pd forms larger clusters on the
$c^+$ surface than on the $c^-$ surface. Correspondingly, we find
that Pd covers a much smaller area of the $c^+$ surface than of
$c^-$. We also find that Pd area is insensitive to deposition rate
at room temperature, as all diffusion and agglomeration processes
are highly activated (for $c^+$) or suppressed (for $c^-$) at this
temperature. On the $c^-$ surface, the area covered by Pd is only
affected when agglomeration plays a role, which occurs when
coverage is above 0.5 ML (Fig.~\ref{f.kmc}(a)). On the $c^+$
surface, Pd coverage area increases as temperature is lowered
(Fig.~\ref{f.kmc}(b)).

The relatively large diffusion barriers on the $c^-$ surface
($\approx$ 0.8~eV) were not sufficient to prevent agglomeration
completely, especially when our simulations were extended until
slightly after deposition was complete. However, we infer that the
clusters on the $c^-$ surface remain much smaller than those on
the $c^+$ surface over a very long time scale (Supplementary Fig.
S3).

Overall, our results support the conclusion of Zhao \emph{et al.}
that Pd cluster sizes are much larger on the $c^+$ surface than on
$c^-$~\cite{Zhao09p1337}. However, Yun \emph{et al.}  also
produced a thorough data set supporting their conclusion that Pd
forms large clusters on both surfaces of
LiNbO$_3$~\cite{Yun09p3145}. One possible explanation is that Yun
and Zhao studied Pd adsorption onto LiNbO$_3$ surfaces with
substantially different compositions.  This could be due to the
presence of different impurities, such as hydroxyl
groups~\cite{lithiumniobate}, which are known to exist on some
LiNbO$_3$ surfaces, or oxygen lattice vacancies~\cite{Yun07p4636}.
Our calculations predict that Pd binding is weaker by 0.6~eV to a
$c^-$ surface terminated with OH instead of OLi, suggesting that
agglomeration would be more favorable with some OH impurities
present. Thus, it is plausible that if there were OH impurities on
the $c^-$ surface studied by Yun \emph{et al.}, they may have
influenced the formation of larger Pd clusters.

In conclusion, we find that the different surface geometries of
oppositely poled LiNbO$_3$ (0001) surfaces lead to substantially
different Pd adsorption geometries. On the $c^-$ surface, strong
Pd bonding to O$'$, the extra surface oxygen present for charge
passivation, leads to larger diffusion and agglomeration barriers.
This lead in turn to less clustering and a more planar geometry
overall on this surface than on $c^+$. In contrast to the
prevalent view that differences in catalyst charging are
responsible for differences in the catalytic properties of metals
adsorbed on polar ferroelectric surfaces, we conclude that the
difference in adsorption geometry predicted here is sufficient to
explain much of the difference in catalytic activity that has been
observed for Pd deposited on oppositely poled LiNbO$_3$
surfaces~\cite{Inoue84p1148}. Because formation of large Pd
clusters is thermodynamically favorable on both $c^+$ and $c^-$,
it is unlikely that the catalytic activity of Pd could be switched
reversibly by switching the polarization of the LiNbO$_3$
substrate. However, the combination of theoretical methods used
here is well suited for the study of other metal/ferroelectric
surface combinations in search of systems with switchable
catalytic activity, and the present example serves as a paradigm
of polarization controlling catalytic cluster geometry and
reactivity.

S. K. was supported by the US Department of Energy through grant
DE-FG02-07ER15920, and A. M. R. by the Air Force Office of
Scientific Research through FA9550-10-1-0248. Computational
support was provided by the High-Performance Computing
Modernization Office of the US Department of Defense.

\bibliography {skim_PdLNO}

\begin{thebibliography}{27}
\expandafter\ifx\csname natexlab\endcsname\relax\def\natexlab#1{#1}\fi
\expandafter\ifx\csname bibnamefont\endcsname\relax
  \def\bibnamefont#1{#1}\fi
\expandafter\ifx\csname bibfnamefont\endcsname\relax
  \def\bibfnamefont#1{#1}\fi
\expandafter\ifx\csname citenamefont\endcsname\relax
  \def\citenamefont#1{#1}\fi
\expandafter\ifx\csname url\endcsname\relax
  \def\url#1{\texttt{#1}}\fi
\expandafter\ifx\csname urlprefix\endcsname\relax\def\urlprefix{URL }\fi
\providecommand{\bibinfo}[2]{#2}
\providecommand{\eprint}[2][]{\url{#2}}

\bibitem[{\citenamefont{Noguera}(2000)}]{Noguera00p367}
\bibinfo{author}{\bibfnamefont{C.}~\bibnamefont{Noguera}}, \bibinfo{journal}{J.
  Phys.: Condens. Matter} \textbf{\bibinfo{volume}{12}}, \bibinfo{pages}{R367 }
  (\bibinfo{year}{2000}).

\bibitem[{\citenamefont{Levchenko and Rappe}(2008)}]{Levchenko08p256101}
\bibinfo{author}{\bibfnamefont{S.~V.} \bibnamefont{Levchenko}}
  \bibnamefont{and} \bibinfo{author}{\bibfnamefont{A.~M.} \bibnamefont{Rappe}},
  \bibinfo{journal}{Phys. Rev. Lett.} \textbf{\bibinfo{volume}{100}},
  \bibinfo{pages}{256101} (\bibinfo{year}{2008}).

\bibitem[{\citenamefont{Yun et~al.}(2007{\natexlab{a}})\citenamefont{Yun, Li,
  Liao, Kampschulte, and Altman}}]{Yun07p4636}
\bibinfo{author}{\bibfnamefont{Y.}~\bibnamefont{Yun}},
  \bibinfo{author}{\bibfnamefont{M.}~\bibnamefont{Li}},
  \bibinfo{author}{\bibfnamefont{D.}~\bibnamefont{Liao}},
  \bibinfo{author}{\bibfnamefont{L.}~\bibnamefont{Kampschulte}},
  \bibnamefont{and} \bibinfo{author}{\bibfnamefont{E.~I.}
  \bibnamefont{Altman}}, \bibinfo{journal}{Surf. Sci.}
  \textbf{\bibinfo{volume}{601}}, \bibinfo{pages}{4636}
  (\bibinfo{year}{2007}{\natexlab{a}}).

\bibitem[{\citenamefont{Inoue et~al.}(1985)\citenamefont{Inoue, Sato, and
  Suzuki}}]{Inoue85p2827}
\bibinfo{author}{\bibfnamefont{Y.}~\bibnamefont{Inoue}},
  \bibinfo{author}{\bibfnamefont{K.}~\bibnamefont{Sato}}, \bibnamefont{and}
  \bibinfo{author}{\bibfnamefont{S.}~\bibnamefont{Suzuki}},
  \bibinfo{journal}{J. Phys. Chem.} \textbf{\bibinfo{volume}{89}},
  \bibinfo{pages}{2827} (\bibinfo{year}{1985}).

\bibitem[{\citenamefont{Park and Baker}(2000)}]{Park00p4418}
\bibinfo{author}{\bibfnamefont{C.}~\bibnamefont{Park}} \bibnamefont{and}
  \bibinfo{author}{\bibfnamefont{R.~T.~K.} \bibnamefont{Baker}},
  \bibinfo{journal}{J. Phys. Chem. B} \textbf{\bibinfo{volume}{104}},
  \bibinfo{pages}{4418} (\bibinfo{year}{2000}).

\bibitem[{\citenamefont{Yun and Altman}(2007)}]{Yun07p15684}
\bibinfo{author}{\bibfnamefont{Y.}~\bibnamefont{Yun}} \bibnamefont{and}
  \bibinfo{author}{\bibfnamefont{E.~I.} \bibnamefont{Altman}},
  \bibinfo{journal}{J. Am. Chem. Soc.} \textbf{\bibinfo{volume}{129}},
  \bibinfo{pages}{15684} (\bibinfo{year}{2007}).

\bibitem[{\citenamefont{Yun et~al.}(2007{\natexlab{b}})\citenamefont{Yun,
  Kampschulte, Li, Liao, and Altman}}]{Yun07p13951}
\bibinfo{author}{\bibfnamefont{Y.}~\bibnamefont{Yun}},
  \bibinfo{author}{\bibfnamefont{L.}~\bibnamefont{Kampschulte}},
  \bibinfo{author}{\bibfnamefont{M.}~\bibnamefont{Li}},
  \bibinfo{author}{\bibfnamefont{D.}~\bibnamefont{Liao}}, \bibnamefont{and}
  \bibinfo{author}{\bibfnamefont{E.~I.} \bibnamefont{Altman}},
  \bibinfo{journal}{J. Phys. Chem. C} \textbf{\bibinfo{volume}{111}},
  \bibinfo{pages}{13951} (\bibinfo{year}{2007}{\natexlab{b}}).

\bibitem[{\citenamefont{Kolpak et~al.}(2007)\citenamefont{Kolpak, Grinberg, and
  Rappe}}]{Kolpak07p166101}
\bibinfo{author}{\bibfnamefont{A.~M.} \bibnamefont{Kolpak}},
  \bibinfo{author}{\bibfnamefont{I.}~\bibnamefont{Grinberg}}, \bibnamefont{and}
  \bibinfo{author}{\bibfnamefont{A.~M.} \bibnamefont{Rappe}},
  \bibinfo{journal}{Phys. Rev. Lett.} \textbf{\bibinfo{volume}{98}},
  \bibinfo{pages}{166101} (\bibinfo{year}{2007}).

\bibitem[{\citenamefont{Li et~al.}(2008)\citenamefont{Li, Zhao, Garra, Kolpak,
  Rappe, Bonnell, and Vohs}}]{Li08p473}
\bibinfo{author}{\bibfnamefont{D.}~\bibnamefont{Li}},
  \bibinfo{author}{\bibfnamefont{M.~H.} \bibnamefont{Zhao}},
  \bibinfo{author}{\bibfnamefont{J.}~\bibnamefont{Garra}},
  \bibinfo{author}{\bibfnamefont{A.}~\bibnamefont{Kolpak}},
  \bibinfo{author}{\bibfnamefont{A.}~\bibnamefont{Rappe}},
  \bibinfo{author}{\bibfnamefont{D.~A.} \bibnamefont{Bonnell}},
  \bibnamefont{and} \bibinfo{author}{\bibfnamefont{J.~M.} \bibnamefont{Vohs}},
  \bibinfo{journal}{Nature Mater.} \textbf{\bibinfo{volume}{7}},
  \bibinfo{pages}{473} (\bibinfo{year}{2008}).

\bibitem[{\citenamefont{Zhao et~al.}(2009)\citenamefont{Zhao, Bonnell, and
  Vohs}}]{Zhao09p1337}
\bibinfo{author}{\bibfnamefont{M.~S.~H.} \bibnamefont{Zhao}},
  \bibinfo{author}{\bibfnamefont{D.~A.} \bibnamefont{Bonnell}},
  \bibnamefont{and} \bibinfo{author}{\bibfnamefont{J.~M.} \bibnamefont{Vohs}},
  \bibinfo{journal}{J. Vac. Sci. Technol.} \textbf{\bibinfo{volume}{27}},
  \bibinfo{pages}{1337} (\bibinfo{year}{2009}).

\bibitem[{\citenamefont{Inoue et~al.}(1984)\citenamefont{Inoue, Yoshioka, and
  Sato}}]{Inoue84p1148}
\bibinfo{author}{\bibfnamefont{Y.}~\bibnamefont{Inoue}},
  \bibinfo{author}{\bibfnamefont{I.}~\bibnamefont{Yoshioka}}, \bibnamefont{and}
  \bibinfo{author}{\bibfnamefont{K.}~\bibnamefont{Sato}}, \bibinfo{journal}{J.
  Phys. Chem.} \textbf{\bibinfo{volume}{88}}, \bibinfo{pages}{1148}
  (\bibinfo{year}{1984}).

\bibitem[{\citenamefont{Inoue et~al.}(1992)\citenamefont{Inoue, Matsukawa, and
  Sato}}]{Inoue92p2222}
\bibinfo{author}{\bibfnamefont{Y.}~\bibnamefont{Inoue}},
  \bibinfo{author}{\bibfnamefont{M.}~\bibnamefont{Matsukawa}},
  \bibnamefont{and} \bibinfo{author}{\bibfnamefont{K.}~\bibnamefont{Sato}},
  \bibinfo{journal}{J. Physs. Chem.} \textbf{\bibinfo{volume}{96}},
  \bibinfo{pages}{2222} (\bibinfo{year}{1992}).

\bibitem[{\citenamefont{Saito et~al.}(2002)\citenamefont{Saito, Yukawa, and
  Inoue}}]{Saito02p10179}
\bibinfo{author}{\bibfnamefont{N.}~\bibnamefont{Saito}},
  \bibinfo{author}{\bibfnamefont{Y.}~\bibnamefont{Yukawa}}, \bibnamefont{and}
  \bibinfo{author}{\bibfnamefont{Y.}~\bibnamefont{Inoue}}, \bibinfo{journal}{J.
  Phys. Chem. B} \textbf{\bibinfo{volume}{106}}, \bibinfo{pages}{10179}
  (\bibinfo{year}{2002}).

\bibitem[{Nor()}]{Norskov08p2163}
\bibinfo{howpublished}{J. K. N{\o}rskov \emph{et al.}, Chem. Soc. Rev.
  \textbf{37}, 2163 (2008)}.

\bibitem[{\citenamefont{Yun et~al.}(2009)\citenamefont{Yun, Pilet, Schwarz, and
  Altman}}]{Yun09p3145}
\bibinfo{author}{\bibfnamefont{Y.}~\bibnamefont{Yun}},
  \bibinfo{author}{\bibfnamefont{N.}~\bibnamefont{Pilet}},
  \bibinfo{author}{\bibfnamefont{U.~D.} \bibnamefont{Schwarz}},
  \bibnamefont{and} \bibinfo{author}{\bibfnamefont{E.~I.}
  \bibnamefont{Altman}}, \bibinfo{journal}{Surf. Sci.}
  \textbf{\bibinfo{volume}{603}}, \bibinfo{pages}{3145} (\bibinfo{year}{2009}).

\bibitem[{\citenamefont{Henkelman et~al.}(2000)\citenamefont{Henkelman,
  Uberuaga, and J\'{o}nsson}}]{Henkelman00p9901}
\bibinfo{author}{\bibfnamefont{G.}~\bibnamefont{Henkelman}},
  \bibinfo{author}{\bibfnamefont{B.~P.} \bibnamefont{Uberuaga}},
  \bibnamefont{and}
  \bibinfo{author}{\bibfnamefont{H.}~\bibnamefont{J\'{o}nsson}},
  \bibinfo{journal}{J. Chem. Phys.} \textbf{\bibinfo{volume}{113}},
  \bibinfo{pages}{9901} (\bibinfo{year}{2000}).

\bibitem[{\citenamefont{Bengtsson}(1999)}]{Bengtsson99p12301}
\bibinfo{author}{\bibfnamefont{L.}~\bibnamefont{Bengtsson}},
  \bibinfo{journal}{Phys. Rev. B} \textbf{\bibinfo{volume}{59}},
  \bibinfo{pages}{12301} (\bibinfo{year}{1999}).

\bibitem[{\citenamefont{Ihm et~al.}(1979)\citenamefont{Ihm, Zunger, and
  Cohen}}]{Ihm79p4409}
\bibinfo{author}{\bibfnamefont{J.}~\bibnamefont{Ihm}},
  \bibinfo{author}{\bibfnamefont{A.}~\bibnamefont{Zunger}}, \bibnamefont{and}
  \bibinfo{author}{\bibfnamefont{M.~L.} \bibnamefont{Cohen}},
  \bibinfo{journal}{J. Phys. C} \textbf{\bibinfo{volume}{12}},
  \bibinfo{pages}{4409} (\bibinfo{year}{1979}).

\bibitem[{\citenamefont{Perdew et~al.}(1996)\citenamefont{Perdew, Burke, and
  Ernzerhof}}]{Perdew96p3865}
\bibinfo{author}{\bibfnamefont{J.~P.} \bibnamefont{Perdew}},
  \bibinfo{author}{\bibfnamefont{K.}~\bibnamefont{Burke}}, \bibnamefont{and}
  \bibinfo{author}{\bibfnamefont{M.}~\bibnamefont{Ernzerhof}},
  \bibinfo{journal}{Phys. Rev. Lett.} \textbf{\bibinfo{volume}{77}},
  \bibinfo{pages}{3865} (\bibinfo{year}{1996}).

\bibitem[{Gia()}]{Giannozzi09p395502}
\bibinfo{howpublished}{P. Giannozzi \emph{et al.}, J. Phys.:Condens. Matter
  \textbf{21}, 395502 (2009)}.

\bibitem[{\citenamefont{Rappe et~al.}(1990)\citenamefont{Rappe, Rabe, Kaxiras,
  and Joannopoulos}}]{Rappe90p1227}
\bibinfo{author}{\bibfnamefont{A.~M.} \bibnamefont{Rappe}},
  \bibinfo{author}{\bibfnamefont{K.~M.} \bibnamefont{Rabe}},
  \bibinfo{author}{\bibfnamefont{E.}~\bibnamefont{Kaxiras}}, \bibnamefont{and}
  \bibinfo{author}{\bibfnamefont{J.~D.} \bibnamefont{Joannopoulos}},
  \bibinfo{journal}{Phys. Rev. B Rapid Comm.} \textbf{\bibinfo{volume}{41}},
  \bibinfo{pages}{1227} (\bibinfo{year}{1990}).

\bibitem[{\citenamefont{Ramer and Rappe}(1999)}]{Ramer99p12471}
\bibinfo{author}{\bibfnamefont{N.~J.} \bibnamefont{Ramer}} \bibnamefont{and}
  \bibinfo{author}{\bibfnamefont{A.~M.} \bibnamefont{Rappe}},
  \bibinfo{journal}{Phys. Rev. B} \textbf{\bibinfo{volume}{59}},
  \bibinfo{pages}{12471} (\bibinfo{year}{1999}).

\bibitem[{Opi()}]{Opium}
\bibinfo{howpublished}{http://opium.sourceforge.net}.

\bibitem[{\citenamefont{Brune}(1998)}]{Brune98p121}
\bibinfo{author}{\bibfnamefont{H.}~\bibnamefont{Brune}},
  \bibinfo{journal}{Surface Science Reports} \textbf{\bibinfo{volume}{31}},
  \bibinfo{pages}{121} (\bibinfo{year}{1998}).

\bibitem[{\citenamefont{Xu et~al.}(2006)\citenamefont{Xu, Henkelman, Campbel,
  and J\'{o}nsson}}]{Xu06p1351}
\bibinfo{author}{\bibfnamefont{L.}~\bibnamefont{Xu}},
  \bibinfo{author}{\bibfnamefont{G.}~\bibnamefont{Henkelman}},
  \bibinfo{author}{\bibfnamefont{C.~T.} \bibnamefont{Campbel}},
  \bibnamefont{and}
  \bibinfo{author}{\bibfnamefont{H.}~\bibnamefont{J\'{o}nsson}},
  \bibinfo{journal}{Surf. Sci.} \textbf{\bibinfo{volume}{600}},
  \bibinfo{pages}{1351 } (\bibinfo{year}{2006}).

\bibitem[{\citenamefont{Xu et~al.}(2007)\citenamefont{Xu, Campbell,
  J\'{o}nsson, and Henkelman}}]{Xu07p3133}
\bibinfo{author}{\bibfnamefont{L.}~\bibnamefont{Xu}},
  \bibinfo{author}{\bibfnamefont{C.~T.} \bibnamefont{Campbell}},
  \bibinfo{author}{\bibfnamefont{H.}~\bibnamefont{J\'{o}nsson}},
  \bibnamefont{and}
  \bibinfo{author}{\bibfnamefont{G.}~\bibnamefont{Henkelman}},
  \bibinfo{journal}{Surf. Sci.} \textbf{\bibinfo{volume}{601}},
  \bibinfo{pages}{3133 } (\bibinfo{year}{2007}).

\bibitem[{\citenamefont{Volk and W\"{o}hlecke}(2008)}]{lithiumniobate}
\bibinfo{author}{\bibfnamefont{T.}~\bibnamefont{Volk}} \bibnamefont{and}
  \bibinfo{author}{\bibfnamefont{M.}~\bibnamefont{W\"{o}hlecke}},
  \emph{\bibinfo{title}{Lithium Niobate: Defects, Photorefraction and
  Ferroelectric Switching}} (\bibinfo{publisher}{Springer},
  \bibinfo{year}{2008}).

\end{thebibliography}


\begin{thebibliography}{10}
\expandafter\ifx\csname natexlab\endcsname\relax\def\natexlab#1{#1}\fi
\expandafter\ifx\csname bibnamefont\endcsname\relax
  \def\bibnamefont#1{#1}\fi
\expandafter\ifx\csname bibfnamefont\endcsname\relax
  \def\bibfnamefont#1{#1}\fi
\expandafter\ifx\csname citenamefont\endcsname\relax
  \def\citenamefont#1{#1}\fi
\expandafter\ifx\csname url\endcsname\relax
  \def\url#1{\texttt{#1}}\fi
\expandafter\ifx\csname urlprefix\endcsname\relax\def\urlprefix{URL }\fi
\providecommand{\bibinfo}[2]{#2}
\providecommand{\eprint}[2][]{\url{#2}}

\bibitem[{\citenamefont{Levchenko and Rappe}(2008)}]{Levchenko08p256101}
\bibinfo{author}{\bibfnamefont{S.~V.} \bibnamefont{Levchenko}}
  \bibnamefont{and} \bibinfo{author}{\bibfnamefont{A.~M.} \bibnamefont{Rappe}},
  \bibinfo{journal}{Phys. Rev. Lett.} \textbf{\bibinfo{volume}{100}},
  \bibinfo{pages}{256101} (\bibinfo{year}{2008}).

\bibitem[{\citenamefont{Saito et~al.}(2004)\citenamefont{Saito, Matsumoto,
  Ohnisi, Akai-Kasaya, Kuwahara, and Aono}}]{Saito04p2057}
\bibinfo{author}{\bibfnamefont{A.}~\bibnamefont{Saito}},
  \bibinfo{author}{\bibfnamefont{H.}~\bibnamefont{Matsumoto}},
  \bibinfo{author}{\bibfnamefont{S.}~\bibnamefont{Ohnisi}},
  \bibinfo{author}{\bibfnamefont{M.}~\bibnamefont{Akai-Kasaya}},
  \bibinfo{author}{\bibfnamefont{Y.}~\bibnamefont{Kuwahara}}, \bibnamefont{and}
  \bibinfo{author}{\bibfnamefont{M.}~\bibnamefont{Aono}},
  \bibinfo{journal}{Jap. J. Appl. Phys} \textbf{\bibinfo{volume}{43}},
  \bibinfo{pages}{2057} (\bibinfo{year}{2004}).

\bibitem[{\citenamefont{Monkhorst and Pack}(1976)}]{Monkhorst76p5188}
\bibinfo{author}{\bibfnamefont{H.~J.} \bibnamefont{Monkhorst}}
  \bibnamefont{and} \bibinfo{author}{\bibfnamefont{J.~D.} \bibnamefont{Pack}},
  \bibinfo{journal}{Phys. Rev. B} \textbf{\bibinfo{volume}{13}},
  \bibinfo{pages}{5188} (\bibinfo{year}{1976}).

\bibitem[{\citenamefont{Ihm et~al.}(1979)\citenamefont{Ihm, Zunger, and
  Cohen}}]{Ihm79p4409}
\bibinfo{author}{\bibfnamefont{J.}~\bibnamefont{Ihm}},
  \bibinfo{author}{\bibfnamefont{A.}~\bibnamefont{Zunger}}, \bibnamefont{and}
  \bibinfo{author}{\bibfnamefont{M.~L.} \bibnamefont{Cohen}},
  \bibinfo{journal}{J. Phys. C} \textbf{\bibinfo{volume}{12}},
  \bibinfo{pages}{4409} (\bibinfo{year}{1979}).

\bibitem[{\citenamefont{Nell and O'Neill}(1996)}]{Nell96p2487}
\bibinfo{author}{\bibfnamefont{J.}~\bibnamefont{Nell}} \bibnamefont{and}
  \bibinfo{author}{\bibfnamefont{H.~S.} \bibnamefont{O'Neill}},
  \bibinfo{journal}{Geochimica et Cosmochimica Acta}
  \textbf{\bibinfo{volume}{60}}, \bibinfo{pages}{2487} (\bibinfo{year}{1996}).

\bibitem[{\citenamefont{Kittel}(1996)}]{Kittel96ISSP}
\bibinfo{author}{\bibfnamefont{C.}~\bibnamefont{Kittel}},
  \emph{\bibinfo{title}{Introduction to Solid State Physics}}
  (\bibinfo{publisher}{John Wiley \& Sons, Inc.}, \bibinfo{year}{1996}),
  \bibinfo{edition}{seventh} ed.

\bibitem[{\citenamefont{Furche}(2001)}]{Furche01p195120}
\bibinfo{author}{\bibfnamefont{F.}~\bibnamefont{Furche}},
  \bibinfo{journal}{Phys. Rev. B} \textbf{\bibinfo{volume}{64}},
  \bibinfo{pages}{195120} (\bibinfo{year}{2001}).

\bibitem[{\citenamefont{Trushin et~al.}(2005)\citenamefont{Trushin, Karim,
  Kara, and Rahman}}]{Trushin05p115401}
\bibinfo{author}{\bibfnamefont{O.}~\bibnamefont{Trushin}},
  \bibinfo{author}{\bibfnamefont{A.}~\bibnamefont{Karim}},
  \bibinfo{author}{\bibfnamefont{A.}~\bibnamefont{Kara}}, \bibnamefont{and}
  \bibinfo{author}{\bibfnamefont{T.~S.} \bibnamefont{Rahman}},
  \bibinfo{journal}{Phys. Rev. B} \textbf{\bibinfo{volume}{72}},
  \bibinfo{pages}{115401} (\bibinfo{year}{2005}).

\bibitem[{\citenamefont{Fan and Gong}(2003)}]{Fan03p117}
\bibinfo{author}{\bibfnamefont{W.}~\bibnamefont{Fan}} \bibnamefont{and}
  \bibinfo{author}{\bibfnamefont{X.}~\bibnamefont{Gong}},
  \bibinfo{journal}{Applied Surface Science} \textbf{\bibinfo{volume}{219}},
  \bibinfo{pages}{117} (\bibinfo{year}{2003}).

\bibitem[{\citenamefont{Zhao et~al.}(2009)\citenamefont{Zhao, Bonnell, and
  Vohs}}]{Zhao09p1337}
\bibinfo{author}{\bibfnamefont{M.~S.~H.} \bibnamefont{Zhao}},
  \bibinfo{author}{\bibfnamefont{D.~A.} \bibnamefont{Bonnell}},
  \bibnamefont{and} \bibinfo{author}{\bibfnamefont{J.~M.} \bibnamefont{Vohs}},
  \bibinfo{journal}{J. Vac. Sci. Technol.} \textbf{\bibinfo{volume}{27}},
  \bibinfo{pages}{1337} (\bibinfo{year}{2009}).

\end{thebibliography}

\end{document}

%\linenumbers
\title{{\Large Supplementary Material:} \\ Polarization-dependence of palladium deposition on
ferroelectric lithium niobate (0001) surfaces}
\author{Seungchul Kim}
\affiliation{The Makineni Theoretical Laboratories, Department of
Chemistry, University of Pennsylvania, Philadelphia, Pennsylvania
19104--6323, USA}
\author{Michael Rutenberg Schoenberg}
\affiliation{The Makineni Theoretical Laboratories, Department of
Chemistry, University of Pennsylvania, Philadelphia, Pennsylvania
19104--6323, USA}
\author{Andrew M. Rappe}
\email[corresponding author:\ ]{rappe@sas.upenn.edu}
\affiliation{The Makineni Theoretical Laboratories, Department of
Chemistry, University of Pennsylvania, Philadelphia, Pennsylvania
19104--6323, USA}
\date{\today }

\maketitle

\makeatletter
\renewcommand{\thefigure}{S\@arabic\c@figure}
\renewcommand{\thetable}{S\@arabic\c@table}
\makeatother

\section{Atomic model of lithium niobate substrates}

\begin{figure}
  \centering
  \includegraphics[angle=0, width=0.45\textwidth]{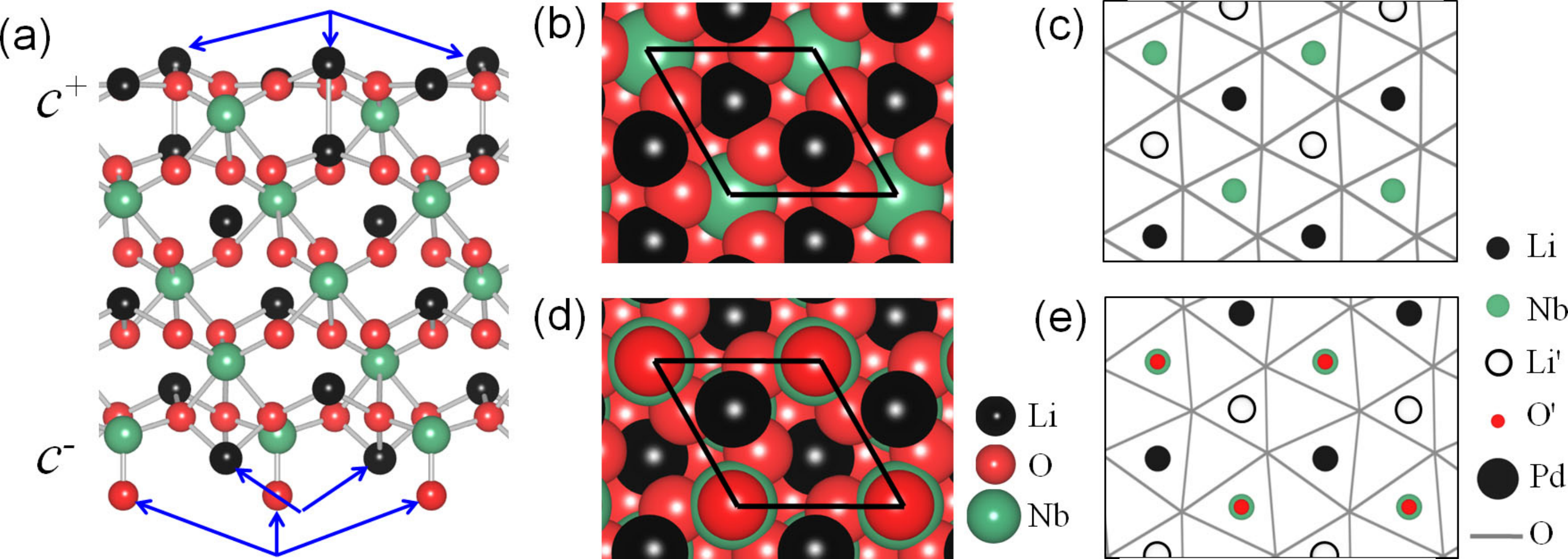}
\caption{Geometry of five-trilayer LiNbO$_3$ slab: (a) side view,
(b,c) top view of $c^+$, and (d,e) top view of $c^-$.  Slabs are
represented using ball and stick (a) and space filling (b,d)
models, or the simplified model used in the main text (c,e). Blue
arrows in (a) denote ions (Li$'$ and O$'$) introduced for surface
charge passivation. Black lines in (b) and (d) indicate the
boundaries of the LiNbO$_3$ primitive surface unit cell. (b) and
(d) are redrawn from Ref.~\cite{Levchenko08p256101}.}
  \label{f.slab}
\end{figure}

The ferroelectric phase of lithium niobate (LiNbO$_3$) has a bulk
structure consisting of layers in a Li-O$_3$-Nb
pattern~\cite{Saito04p2057}.  Here we show LiNbO$_3$ with its
thermodynamically preferred surface terminations, as predicted by
Levchenko and Rappe (Fig.\ S1).  Li atoms are added to the
positive ($c^+$) surface, and both Li and O atoms are added to the
negative surface ($c^-$) of LiNbO$_3$ for charge passivation.
Passivation atoms are denoted Li$'$ and O$'$.  These surface
terminations make the overall stoichiometry of five trilayer slab
used in this study, from $c^+$ (left) to $c^-$ (right),
Li$'$-(Li-O$_3$-Nb)$_5$-Li$'$O$'$.

\section{Supplementary DFT methods}

In all density functional theory (DFT) calculations, we used
3~$\times$~3~$\times$~1 $k$-point grids~\cite{Monkhorst76p5188},
plane wave energy cutoffs of 50~Ry~\cite{Ihm79p4409}, and force
tolerances of 0.01~eV/\AA.  When necessary, we included spin as a
degree of freedom in our calculations.  We checked the convergence
of our monomer and dimer adsorption energies calculated in
$\sqrt{3}\times\sqrt{3}$ supercells on five layer thick slabs by
using either $\sqrt{7}\times\sqrt{7}$ supercells or seven trilayer
thick slabs.  The resulting energies agreed within 0.1~eV/Pd.

\section{Adsorption Energies}
\begin{table}[t]
  %\begin{ruledtabular}
  \begin{tabular}{c|cc|ccc|c}
  \hline
  \hline
     \# of Pd  & \multicolumn{2}{c|} {positive ($c^+$)} & \multicolumn{3}{c|}{ negative ($c^-$)} & free Pd$_n$ \\
             \cline{2-6}
     atoms    & $E^{a}_{ads}$& $E^{c}_{ads}$ & $E^{a}_{ads}$ & $E^{c}_{ads}$ & $E^{\rm Pd-X}_{\rm bond}$ & $E^{\rm Pd-Pd}_{\rm bond}$ \\
   \hline
   1 & 0.95  &  --   & 2.02  &  --   & 1.01 & --   \\
   2 & 1.20  & 0.62  & 2.01  & 1.44  & 1.01 & 1.15 \\
   3 & 1.56  & 0.44  & 2.27  & 1.15  & 1.13 & 1.12 \\
   4 & 1.84  & 0.26  & 2.34  & 0.77  & 1.02 & 1.05 \\
   5 &  --   &  --   & 2.35  & 0.65  & 0.98 & 0.94 \\
  \hline
  \hline
  \end{tabular}
  %\end{ruledtabular}
\caption{Per-atom adsorption energies of Pd on LiNbO$_3$ (LNO),
relative to infinitely separated Pd atoms ($E^{a}_{ads}$) and free
Pd clusters ($E^{c}_{ads}$). We define $E^{a}_{ads} = (E_{\rm
Pd-LNO} - E_{\rm LNO} -nE_{\rm Pd})/n$ and $E^{c}_{ads} = (E_{\rm
Pd-LNO} - E_{\rm LNO} - E_{{\rm Pd}_{n}})/n$,  where $n$ is the
number of Pd atoms. On the $c^-$ surface, we report average Pd-X
bond energies, $E^{\rm Pd-X}_{\rm bond}$= $n$$E^{a}_{ads}$/$N_{\rm
bond}$, where $N_{\rm bond}$ is the number of Pd-X bonds.  We also
report average Pd-Pd bond energies, $E^{\rm Pd-Pd}_{\rm bond}$, in
free Pd clusters for comparison. All energies are given in eV.}
  \label{t.Ead}
\end{table}

In our analysis of the energetics of Pd adsorption on LiNbO$_3$,
we consider two quantities for per-atom adsorption energy,
$E^{a}_{ads}$ and $E^{c}_{ads}$, defined as the difference between
the energy of Pd-LNO and the sum of the energies of the LiNbO$_3$
slab and either infinitely separated ($E^{a}_{ads}$) or clustered
($E^{c}_{ads}$) Pd atoms  divided by the number of Pd atoms
(Table~\ref{t.Ead}). We note that on the $c^+$ surface,
$E^{c}_{ads}$ becomes very small as the number of adsorbed Pd
atoms increases, because of a mismatch between ideal Pd-Pd bond
lengths and distances between Pd binding sites, and because most
of the cluster not in contact with LNO. In contrast, $E^{c}_{ads}$
remains relatively high even for pentamers on $c^-$, largely due
to Pd-O$'$ bonding. In the main text, we explained many of our
conclusions about energetics on the $c^-$ surface in terms of the
total number of Pd-Pd and Pd-O$'$ bonds (hereafter denoted Pd-X
bonds).  This analysis was justified by the fact that average Pd-X
bond energies ($E^{\rm Pd-X}_{\rm bond}$), defined as
$nE^{a}_{ads}$ divided by the number of Pd-X bonds ($N_{\rm
bond}$), were similar for all cluster sizes studied and were also
similar to average bond energies in free Pd clusters ($E^{c}_{\rm
bond}$).  We show explicit values for $E^{\rm Pd-Pd}_{\rm bond}$
and $E^{\rm Pd-X}_{\rm bond}$ in Table~\ref{t.Ead}.

Since Pd-O$'$ bonding makes the key contribution to the difference
in adsorption behavior between the $c^+$ and $c^-$ surfaces, we
wanted to be sure that DFT-GGA does not misestimate the
favorability of Pd-O interactions.  To do this, we considered the
case of bulk PdO.  We compared our DFT-GGA energy of formation of
PdO from individual Pd and O atoms, defined as $E^{\rm atom}_{\rm
O}+E^{\rm atom}_{\rm Pd}-E_{\rm PdO}$, to the sum of the
experimental values for  PdO formation energy~\cite{Nell96p2487},
Pd bulk cohesive energy~\cite{Kittel96ISSP}, and half of the O$_2$
bonding energy~\cite{Furche01p195120}.  Our DFT-calculated value
was only 2.1\% greater than the experimental value, suggesting
that DFT predicts Pd-O interactions well and that the strong Pd-O
interactions seen on the $c^-$ surface are valid.  As a note, we
did not use the DFT PdO formation energy as a metric for
comparison, because DFT-GGA is known to predict the energy of
O$_2$ quite poorly~\cite{Furche01p195120}.

\section{Kinetic Monte Carlo simulation details}

\begin{figure}
 \centering
   \includegraphics[angle=0, width=0.35\textwidth]{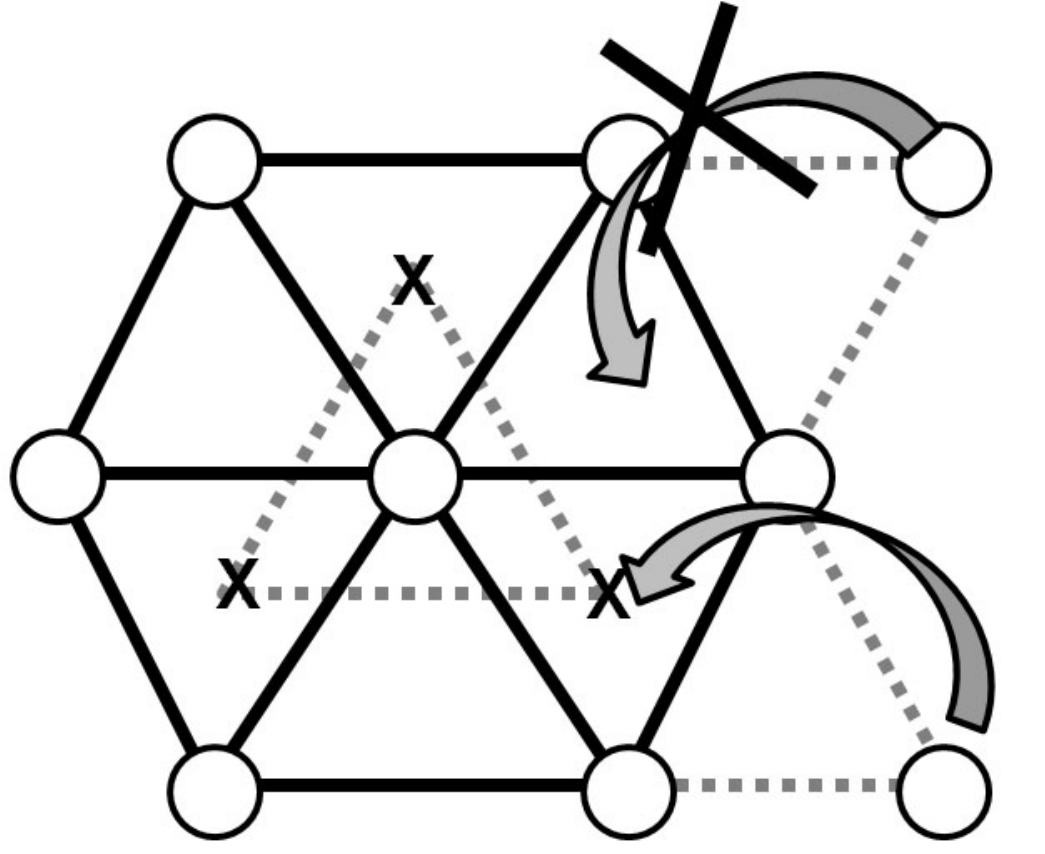}
\caption{ An FCC lattice is not sufficient to allow all possible
agglomeration events to occur.  Of the two agglomeration processes
pictured, one is allowed by the FCC lattice, while the other,
denoted by a red 'X', would also require HCP lattice points on the
second layer of the simulation lattice.  We thus used an
A--(ABC)--(ABC)$\cdots$ lattice as described in the text. Open
circles denote Pd atoms and `X's denote points on the second layer
of the simulation lattice.}
 \label{f.lattice}
\end{figure}

Our kinetic Monte Carlo (KMC) simulations were performed in
periodic 10~nm~$\times$~10~nm cells.  We included 15 layers of Pd
binding sites above the surface, though no Pd atoms reached the
top layer of the cell in any of our simulations.  All Pd binding
sites on the layer in contact with the LiNbO$_3$ surface were
mapped onto their corresponding points on the triangular oxygen
lattice. Beginning with the second layer above the surface, we
used three sublattices to allow all possible Pd cluster
configurations such that the lattice is
substrate--A--(ABC)--(ABC)$\cdots$.  The use of this lattice,
which is much denser than the face-centered cubic (FCC) lattice of
bulk Pd, was necessary to describe some diffusion events that can
occur for small numbers of Pd atoms, but is forbidden by the FCC
lattice.  For example, half of all possible agglomeration
processes are forbidden using an FCC lattice, because lattice
points are present that correspond to only half of the surface's
hollow sites (Fig.~\ref{f.lattice}).  Additionally, hopping of Pd
atoms between FCC and hexagonal close-packed (HCP) hollow sites on
(111) cluster facets is forbidden by a pure FCC lattice. To
compensate for the fact that some points in the lattice used in
our simulation are closer together than a reasonable Pd-Pd bond
distance, we imposed two rules.  First, the same site cannot be
occupied on adjacent layers of the lattice.  Additionally, within
the same layer, adjacent sites on different sublattices cannot be
occupied.

We built our simulations primarily based on our DFT calculations
of the activation barriers of diffusion and agglomeration of up to
four atoms. On the $c^+$ surface, it was unnecessary to consider
diffusion of larger clusters, because our deposition time interval
was sufficiently long that we observed monomers to aggregate into
clusters before other monomers are deposited in the same vicinity.
Diffusion of large clusters on the $c^-$ surface could also be
neglected, because they require the breaking of two or more Pd-X
bonds and thus have activation barriers sufficient to prevent
their occurrence on the time scale that we considered. As
described below, in cases where we had not calculated activation
barriers directly, such as monoatomic motion on cluster facets, we
interpolated reasonable barriers.

We permitted diffusion of Pd atoms that were directly in contact
with the LiNbO$_3$ surface or on cluster facets.  On the LiNbO$_3$
surface, we permitted the following diffusion processes:  1)
nearest-neighbor monomer hopping, 2) agglomeration/deagglomeration
(movement of Pd atom from layer 1 (2) onto layer 2 (1)), 3) dimer
walking and sliding on $c^+$, 4) trimer walking and sliding on
$c^+$, 5) trimer flipping on $c^-$, 6) tetramer rolling and 7)
dissociation (some of these processes are pictured in Fig.~2 in
the main text). On cluster facets, we permitted monomer hopping
only.  Though concerted movements of multiple atoms can occur on
metal cluster facets, these generally affect cluster morphology,
but not cluster size (number of atoms present).  Since our primary
interest is only in seeing differences in cluster size, this
simplification of Pd behavior on cluster facets is acceptable.

\begin{figure*}[!htb]
  \centering
  \includegraphics[angle=0, width=0.8\textwidth]{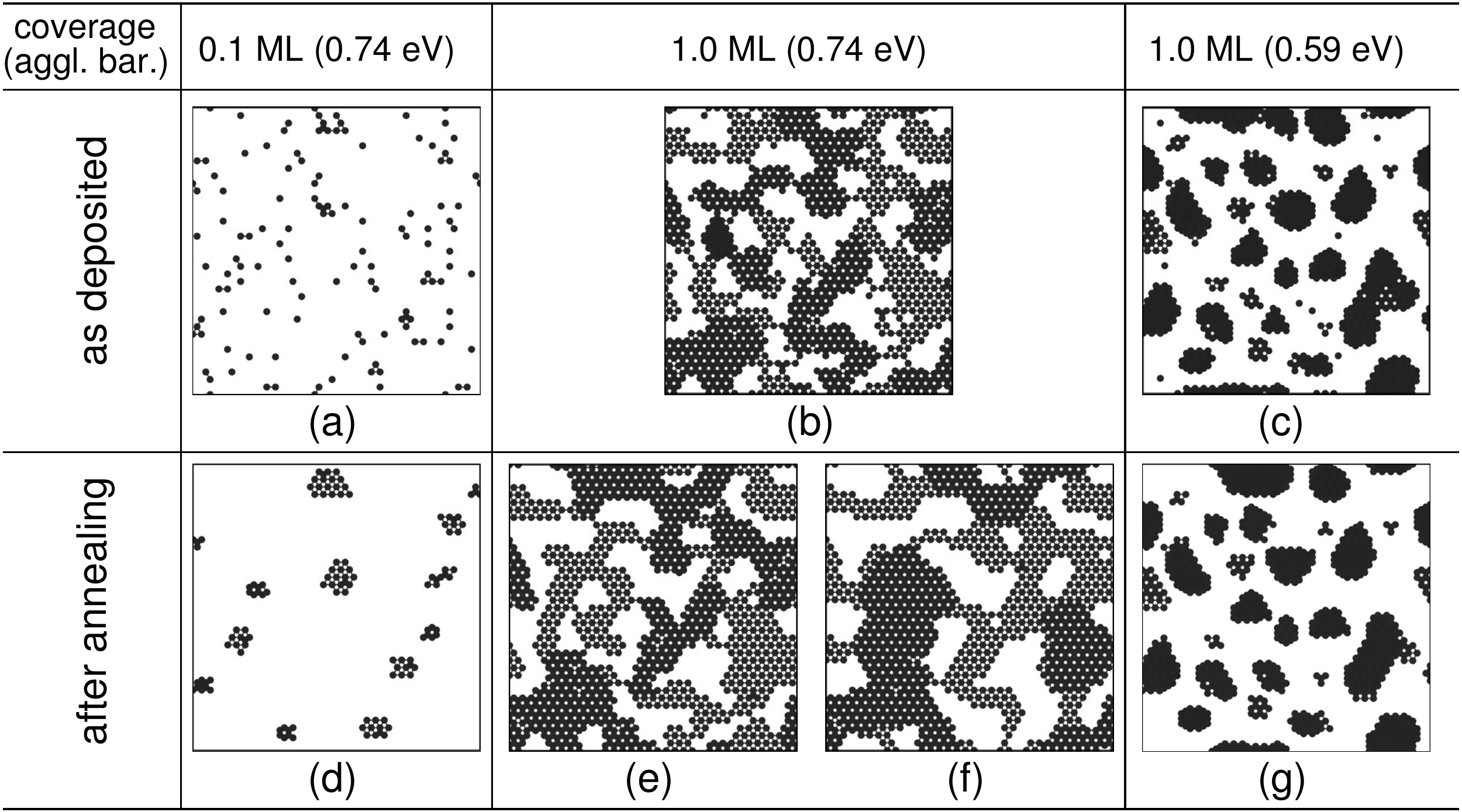}
  \caption{Pd adsorption geometries predicted by kinetic Monte
Carlo simulations on the $c^-$ surface of LiNbO$_3$ immediately
after deposition ends (a-c) and 2-15 minutes later (d-e).  These
simulations were conducted at different coverages (0.1~ML and
1.0~ML), different temperatures (300 and 400~K), and with the two
different agglomeration barriers predicted by DFT (0.59~eV and
0.74~eV).  All simulations were conducted at a deposition rate of
0.025~ML/s.  Times waited after the completion of deposition were:
(d) 15 minutes, (e) 10 minutes, (f) 2 minutes, and (g) 3 minutes.
All simulations are conducted at 300~K, except for (f), which is
at 400~K.}
 \label{f.ann}
\end{figure*}

It is well known that the activation barriers for monomeric
diffusion on the (111) surfaces of FCC metals are quite low
relative to the bulk metal-metal bond energy. Our nudged elastic
band (NEB) calculations show that the activation barrier for
monoatomic hopping from from FCC to HCP hollow sites on (111)
surface of palladium is 0.11~eV, despite the fact that the Pd-Pd
bulk bond energy is nearly six times larger (0.59~eV). This is
explained by the fact that during monomer hopping between
nearest-neighbor sites, new bonds start to form before others are
completely broken.  In our KMC simulations, we interpolated
reasonable activation barriers for diffusion events on Pd cluster
facets using insights from Trunshin and coworker's studies of
nearest-neighbor monoatomic hopping on copper surfaces and
steps~\cite{Trushin05p115401}.  By compensating for the difference
in cohesive energy between bulk Pd and Cu, we estimated that the
activation barrier for nearest neighbor monomer hopping is 0.2~eV
per Pd-Pd bond broken during transition state formation.  We also
showed that the qualitative results of some of our simulations
were insensitive to the use of a 0.3~eV barrier for this process.
In next nearest neighbor monomer hopping, which is analogous to
the agglomeration processes pictured in Fig.~2 but generalized to
movements between any two levels in the simulation lattice
(starting with level 2), Pd-Pd bonds are broken completely during
transition state formation.  Thus, we estimated the activation
barrier for this process based on the cost of Pd-Pd bond breaking.
The cost of breaking a bond in a Pd$_{20}$ cluster cut from bulk
Pd is 0.52~eV.  Similarly the cost of dimer dissociation is
0.55~eV.  Thus, for next-nearest neighbor hopping on Pd cluster
facets, we used a barrier of  0.5~eV per Pd-Pd bond broken.

\section{Aggregation and agglomeration on $c^-$ after deposition}

The relatively large diffusion barriers on the c$^-$ surface
($\approx$0.8~eV) were not sufficiently high to prevent
agglomeration completely, especially when our simulations were
extended until slightly after deposition was complete. We have
conducted KMC simulations during deposition and after annealing
for 2-15~minutes for both low (0.1~ML) and high (1.0~ML) coverages
and at different temperatures (300 or 400~K).  We also conducted
our simulations with agglomeration barriers (0.59 and 0.74~eV) for
the two different possible cluster agglomeration processes
predicted by DFT on the $c^-$ surface (See Fig. 2(g,h) and Table I
in the main text).  These data are reported in Fig.~\ref{f.ann}.

At low coverage (0.1~ML),  Pd atoms are highly dispersed at the
end of deposition (Fig.~\ref{f.ann}(a)).  However, within 15
minutes, all but two Pd clusters are immobile, suggesting that on
a slightly longer time scale, all Pd atoms will join immobile
clusters (Fig.~\ref{f.ann}(d)). However, these clusters are quite
small compared to the single cluster that forms on the $c^+$
surface under the same conditions. Two factors contribute to the
formation of smaller clusters on the $c^-$ surface than on $c^+$:
lower maximum mobile cluster size, and the higher number density
of clusters on the surface during deposition. On the $c^+$
surface, the mismatch between the distances between monomeric Pd
binding sites and Pd-Pd equilibrium bond distances makes the
activation barrier for motion of relatively large clusters rather
low.  For example, our predicted barrier for tetramer rolling was
similar to that for monomer hopping, largely because this process
requires movement of only one Pd atom out of a monomeric potential
well. Other work suggests that this condition may be met by Pd
clusters as large as 13 atoms~\cite{Fan03p117}, and thus, that
clusters of this size may be mobile on the $c^+$ surface.  In
contrast, the diffusion of clusters larger than a tetramer on the
$c^-$ surface requires breakage of at least two Pd-X bonds, making
these diffusion events extremely unlikely on the time scale of
deposition.  A second factor contributing to the formation of
large clusters on the $c^+$ surface is the fact that, because
barriers for diffusion are so low, many diffusion events can occur
during the average waiting time between addition of new atoms to
the simulation cell on this surface.  This means that only a few
atoms at a time will be separate from the large immobile cluster,
greatly reducing the probability that a second immobile cluster
will form.  The opposite is true on the $c^-$ surface, where
larger diffusion barriers mean that on average multiple Pd atoms
are deposited inside the simulation cell between diffusion events
of a single atom or cluster.  As a result, many individual
clusters are present in any given area of the simulation cell,
promoting the formation of many small immobile clusters.

At high coverage (1.0~ML),  Pd atoms continue to rearrange after
deposition is complete until all atoms make at least three Pd-X
bonds (Fig.~\ref{f.ann}(b,e)).  Once this occurs, agglomeration
can only occur on an extremely long time scale, because it
requires breakage of multiple Pd-X bonds.  Thus, the Pd coverage
area of these clusters will remain relatively constant.  This was
true even when our surface was annealed at a higher temperature
(400~K, Fig.~\ref{f.ann}(f)), consistent with the finding of Zhao
and colleagues that the Pd geometry on the $c^-$ surface is stable
up to 425~K~\cite{Zhao09p1337}.

Even when we used the lower  of our two agglomeration barriers
predicted from DFT on the $c^-$ surface (Fig. 2(g) in the main
text), multiple distinct clusters formed at 1 ML coverage
(Fig.~\ref{f.ann}(c,g)). At low coverage we observed results
similar to those for the higher agglomeration barrier
(Fig.~\ref{f.ann}(d)), but with taller individual clusters.
Overall, our simulations with this lower agglomeration barrier on
the $c^-$ surface predicted Pd coverage areas that were
substantially higher than those on the $c^+$ surface, but lower
than the simulations on the $c^-$ surface with the higher
agglomeration barrier.

In sum, our DFT and KMC  calculations predict that though Pd forms
slightly three dimensional clusters on the $c^-$ surface of
LiNbO$_3$, its overall adsorption pattern on this surface is much
more dispersed than that on the $c^+$ surface.

\bibliography {skim_PdLNO}